\documentclass[apj]{emulateapj}
\usepackage{apjfonts}

\newcommand{\eg}{e.g., }
\newcommand{\ie}{i.e., }
\newcommand{\Msun}{M_{\odot}}
\newcommand{\kms}{km~s$^{-1}$}
\newcommand{\ergs}{erg~s$^{-1}$}
\newcommand{\Fefs}{$^{56}$Fe}
\newcommand{\Cofs}{$^{56}$Co}
\newcommand{\Nifs}{$^{56}$Ni}

\newcommand{\KE}{E_{\rm K}}
\newcommand{\vph}{v_{\rm ph}}
\newcommand{\Tph}{T_{\rm ph}}
\def\gsim{\mathrel{\rlap{\lower 4pt \hbox{\hskip 1pt $\sim$}}\raise 1pt
\hbox {$>$}}}
\def\lsim{\mathrel{\rlap{\lower 4pt \hbox{\hskip 1pt $\sim$}}\raise 1pt
\hbox {$<$}}}
\def\ion#1#2{{\rm #1}~{\sc #2}}

\shorttitle{The Outermost Ejecta of Type Ia Supernovae}
\shortauthors{Tanaka et al.}

\begin{document}

\title{The Outermost Ejecta of Type I\lowercase{a} Supernovae}
\author{
Masaomi Tanaka\altaffilmark{1},
Paolo A. Mazzali\altaffilmark{2,3,4},
Stefano Benetti\altaffilmark{5},
Ken'ichi Nomoto\altaffilmark{1,4,6},
Nancy Elias-Rosa\altaffilmark{2},
Rubina Kotak\altaffilmark{7},
Giuliano Pignata\altaffilmark{8,9},
Vallery Stanishev\altaffilmark{10}, and
Stephan Hachinger\altaffilmark{2}
}

\altaffiltext{1}{Department of Astronomy, Graduate School of Science, University of Tokyo, Hongo 7-3-1, Bunkyo-ku, Tokyo 113-0003, Japan; mtanaka@astron.s.u-tokyo.ac.jp}
\altaffiltext{2}{Max-Planck-Institut f\"{u}r Astrophysik, Karl-Schwarzschild-Str. 1, D-85741 Garching bei M\"{u}nchen, Germany}
\altaffiltext{3}{Istituto Nazionale di Astrofisica - Osservatorio Astronomico di Triste, Via Tiepolo 11, I-34131 Triste, Italy}
\altaffiltext{4}{Research Center for the Early universe, School of Science, University of Tokyo, Bunkyo-ku, Tokyo 113-0033, Japan}
\altaffiltext{5}{Istituto Nazionale di Astrofisica - Osservatorio Astronomico di Padova, vicolo dell'Osservatorio 5 - 35122 Padova, Italy}
\altaffiltext{6}{Institute for the Physics and Mathematics of the Universe, 
University of Tokyo, Kashiwa, Chiba 277-8582, Japan}
\altaffiltext{7}{Astrophysics Research Centre, School of Mathematics and Physics, Queen's University Belfast, BT7 1NN, UK}
\altaffiltext{8}{Departamento de Astronom\'ia y Astrof\'isica, Pontificia Universidad Cat\'olica de Chile, Casilla 306, Santiago 22, Chile}
\altaffiltext{9}{Departamento de Astronom\'ia, Universidad de Chile, Casilla 36-D, Santiago, Chile}
\altaffiltext{10}{Department of Physics, Stockholm University, AlbaNova University Center, SE-10691 Stockholm, Sweden}

\begin{abstract}
The properties of the highest velocity ejecta of normal Type Ia supernovae 
(SNe Ia) are studied via models of very early optical spectra of 6 SNe.
At epochs earlier than 1 week before maximum, SNe with a rapidly evolving 
\ion{Si}{ii} $\lambda$6355 line velocity (HVG) have a larger photospheric 
velocity than SNe with a slowly evolving \ion{Si}{ii} $\lambda$6355 line 
velocity (LVG). 
Since the two groups have comparable luminosities, 
the temperature at the photosphere is higher in LVG SNe. 
This explains the different overall spectral appearance of HVG and LVG SNe.
However, the variation of the \ion{Ca}{ii} and \ion{Si}{ii} absorptions at the 
highest velocities ($v \gsim 20,000$ \kms) suggests that additional factors, 
such as asphericity or different abundances in the progenitor white dwarf, 
affect the outermost layers.
The \ion{C}{ii} $\lambda$6578 line is marginally detected in 3 LVG SNe,  
suggesting that LVG undergo less intense burning.
The carbon mass fraction is small, only less than 0.01 near the photosphere, 
so that he mass of unburned C is only $\lsim 0.01 \Msun$.
Radioactive \Nifs\ and stable Fe are detected in both LVG and HVG SNe.
Different Fe-group abundances in the outer layers may be one of the reasons 
for spectral diversity among SNe Ia at the earliest times.
The diversity among SNe Ia at the earliest phases could also indicate an 
intrinsic dispersion in the width-luminosity relation of the light curve.
\end{abstract}


\section{Introduction}
\label{sec:intro}

Type Ia supernovae (SNe Ia) are believed to be thermonuclear explosions 
of CO white dwarfs (WDs).
Because of their high luminosity together with a tight
relation between the maximum brightness and decline rate of the light curve
(LC), SNe Ia are one of the most accurate distance indicators 
(\eg Phillips 1993; Riess et al. 1998; Perlmutter et al. 1999; 
Knop et al. 2003; Riess et al. 2004; 
Astier et al. 2006; Wood-Vasey et al. 2007).

In addition to the homogeneity of the LC, 
optical spectra of SNe Ia are also rather homogeneous. 
The ratio of two \ion{Si}{ii} lines in spectra at maximum
correlates well with the SN luminosity (Nugent et al. 1995).
The width of the Fe emission line in the nebular spectra 
also correlates with the LC decline rate (Mazzali et al. 1998).
These works show that the spectral properties both at maximum brightness 
and at the nebular phases are determined primarily by the maximum luminosity.

However, recent studies using large samples revealed large diversity in the 
early spectra of SNe Ia, in particular in the line velocities (see \eg Hatano
et al. 2000; Benetti et al. 2005, hereafter B05; Branch et al. 2006). B05
divided SNe Ia into three groups according to the post-maximum evolution of the
\ion{Si}{ii} $\lambda$6355 line velocity.  High Velocity Gradient (HVG) SNe
have a large \ion{Si}{ii} line velocity that decreases rapidly after maximum.
On the other hand, Low Velocity Gradient (LVG) SNe have a lower  velocity with
a slower post-maximum evolution (see Fig. 1 in B05).  HVG and LVG SNe have
similar maximum luminosities, implying that line velocity is independent of
luminosity or LC shape (see Fig. 3 in Hachinger, Mazzali \& Benetti 2006).
FAINT SNe are characterized by a lower luminosity and a low \ion{Si}{ii} line
velocity, whose evolution is comparable to that of LVG SNe. 

The origin of the spectral diversity is not yet fully understood, reflecting
uncertainties on the explosion mechanism. Despite general consensus that 
burning starts as a subsonic deflagration (Nomoto, Sugimoto \& Neo 1976;
Nomoto, Thielemann \& Yokoi 1984), it is still unclear whether or not a
transition to a supersonic detonation (Khokhlov 1991b) occurs. 
In recent years, the explosion mechanism has been studied by 
detailed numerical simulations free from spherically symmetry 
(\eg Reinecke,
Hillebrandt, \& Niemeyer 2002; Gamezo et al. 2003; Plewa, Calder \& Lamb 2004;
R\"opke \& Hillebrandt 2005; Gamezo, Khokhlov \& Oran 2005). 

Early phase spectra are an unique tool to investigate the state of the outer
layers of the ejecta, \ie the final fate of the burning front.  Furthermore, as
diversity in line velocities is largest at pre-maximum phases, modeling spectra
at these epochs is likely to yield clues as to the origins of this diversity.
Although several studies focussed on spectral diversity at maximum light (e.g.
Bongard et al. 2006; Branch et al. 2006; Hachinger et al. 2006), only few have
addressed pre-maximum spectra (e.g. Mazzali et al. 1993, Fisher et al. 1997;
Mazzali 2001; Branch et al. 2007).  This is mostly due to the lack of very
early data, which stems from the difficulty in discovering SNe well before
maximum.  In order to remedy this situation, the European Supernova
Collaboration (ESC \footnote{http://www.mpa-garching.mpg.de/\~{}rtn/}) has
endeavored to gather very early spectra of nearby SNe Ia.

One of the most interesting results of very early SN Ia spectroscopy is the
possible presence of unburned carbon in the outer layers. 
This may be a critical signature of the outer extent of burning. 
The \ion{C}{ii} $\lambda$6578 line has been detected in SN 1998aq 
(at $\sim$ 11,000 \kms, Branch et al. 2003) and in SN 1999ac  
(at $\sim$ 16,000 \kms, Garavini et al. 2005).  
The clearest \ion{C}{ii} features were seen in a pre-maximum spectrum of 
SN 2006D (at $\sim$ 12,000 \kms, Thomas et al. 2006)
\footnote{In SN 2006gz, the \ion{C}{ii} line is clearly seen at 
$\sim$ 15,000 \kms, although the SN is not normal and
possibly has a super-Chandrasekhar mass progenitor (Hicken et al. 2007),
as suggested for SN 2003fg (Howell et al. 2006)}.
On the other hand, \ion{C}{i} in near-infrared (NIR) spectra has never 
been firmly detected (Marion et al. 2006).

High-velocity absorption features (HVFs, $v \gsim 20,000$ \kms) in the 
\ion{Ca}{ii} IR triplet (\eg Hatano et al. 1999; Mazzali et al. 2005a;  
Thomas et al. 2004; Quimby et al. 2006) may be a feature of all SNe Ia at the 
earliest epochs (Mazzali et al. 2005b), and may indicate aspherical ejecta, as 
suggested by the high polarization level 
(Wang et al. 2003; Kasen et al. 2003).  
They may result from the explosion itself (Thomas et al. 2004; Mattila et al. 
2005) or from the interaction of the ejecta with circumstellar medium 
(CSM; Gerardy et al. 2004; Mazzali et al. 2005a). 
The variation and distribution of the strength of \ion{Ca}{ii} HVFs may be 
explained by line-of-sight effects if HVFs result from aspherical structures 
like a few blobs or a thick disk at $v \sim 18-25,000$ \kms\ 
(Tanaka et al. 2006).

In this paper, we use SNe Ia pre-maximum optical spectra to study the 
physical properties of the ejecta,
modeling the spectra using a numerical code.
Details of the method are presented in \S \ref{sec:method}.
The results of the modeling are shown in \S \ref{sec:results}.
In \S \ref{sec:discussion}, our results are summarized
and the implications are discussed. 
Finally, conclusions are in \S \ref{sec:conclusions}.

\section{METHOD OF ANALYSIS}
\label{sec:method}

Spectra of 6 nearby SNe Ia observed earlier than 1 week before maximum 
are studied (Table \ref{tab:prop}).
The data were obtained by the ESC campaign.
The ESC has collected photometric and spectroscopic data of a number of 
nearby SNe Ia with very good temporal and wavelength coverage.
SNe with very early spectra include
SN 2001el (LVG; Mattila et al. 2005), 
SN 2002bo (HVG; Benetti et al. 2004), 
SN 2002dj (HVG; Pignata et al. 2005), 
SN 2002er (HVG; Pignata et al. 2004; Kotak et al. 2004),
SN 2003cg (LVG; Elias-Rosa et al. 2006),
and SN 2003du (LVG; Stanishev et al. 2007).
Our sample does not include FAINT SNe.

To investigate the physical conditions of the ejecta, such as temperatures,
ionization profiles, and element abundance distribution, 
we use a Monte Carlo spectrum
synthesis code (Mazzali \& Lucy 1993), which was used for several SNe  (\eg
Mazzali et al. 1993).  The code requires as input the bolometric luminosity
($L$), photospheric velocity ($\vph$), and a model of the density and abundance
distribution in the SN ejecta.

The code assumes a sharply defined spherical photosphere.
The position of the photosphere is expressed in terms of a velocity 
thanks to the homologous expansion ($v \propto r$).
The temperature structure of the optically thin atmosphere is solved in 
radiative equilibrium tracing a large number of energy packets as they 
propagate in the SN ejecta.

Starting from a trial temperature distribution (radiation temperature, $T_R$),
the population of the excited level ($n_j$; $j=1$ for the ground state) is 
computed as
\begin{equation}
\frac{n_j}{n_1} = W \frac{g_j}{g_1} e^{-\epsilon_j /k_B T_R},
\label{eq:level}
\end{equation}
where $g_j$ and $\epsilon_j$ are the statistical weight  
and the excitation energy from the ground level, respectively.
Here W is the so called dilution factor, which is defined as
\begin{equation}
J = WB(T_R).
\end{equation}
In Eq\. (1), W is set to unity for metastable levels (Lucy 1999).
The ionization regime is computed using a modified nebular approximation 
(Abbott \& Lucy 1985, Mazzali \& Lucy 1993):
\begin{equation}
\frac{N_{i+1}N_e}{N_i} 
= \eta W \left( \frac{T_e}{T_R} \right) 
 \left( \frac{N_{i+1}N_e}{N_i}  \right)^{*}_{T_R},
\label{eq:ion}
\end{equation}
where $N_e$ is the electron density and $T_e$ the electron temperature.
The starred term on the right hand side is the value computed using the Saha 
equation with the temperature $T_R$; the partition functions entering the Saha 
equation are, however, computed with Eq. (\ref{eq:level}).
In Eq\. (\ref{eq:ion}), $\eta$ is defined as
\begin{equation}
\eta = \delta \zeta + W(1-\zeta),
\end{equation}
where $\zeta$ is the fraction of recombinations going directly to the ground 
state. Mazzali \& Lucy (1993) introduced a correction factor 
$\delta$ for an optically thick continuum at shorter wavelengths 
than the \ion{Ca}{ii} ionization edge 
(for a detailed definition, see Eqs. (15) and (20) in Mazzali \& Lucy 1993).
The second term in Eq\. (4) represents ionization from excited levels
(Mazzali \& Lucy 1993).

The Monte Carlo radiation packets are followed through the atmosphere, 
where they can undergo electron scattering or interact with spectral lines.
For line scattering, Sobolev approximation,
which is a sound approximation in a rapidly expanding medium, is applied.
From the level populations, a Sobolev line opacity is obtained: 
\begin{equation}
\tau_{lu} = \frac{hct}{4\pi} (B_{lu}n_l - B_{ul}n_u).
\end{equation}
Here $t$ is the time since the explosion and $B_{lu}$ and $B_{ul}$ 
are Einstein $B$-coefficients.
The effect of photon branching is considered
(Lucy 1999; Mazzali 2000; see also Pinto \& Eastman 2000).

The Monte Carlo experiment gives a flux in each radial point, and 
then a frequency moment
\begin{equation}
\bar{\nu} = \frac{\int \nu J_{\nu} d\nu}{\int J_{\nu} d\nu}
\end{equation}
is computed.
Hence the temperature structure can be determined via
\begin{equation}
\bar{x} = \frac{h\bar{\nu}}{k_B T_R}.
\end{equation}
Here $\bar{x}$ represents for mean energy of blackbody radiation
($\bar{x} = 3.832$).
The electron temperature is crudely assumed to be $T_e = 0.9 T_R$ to simulate 
approximately\footnote{This assumption does not affect the spectral fitting, 
and thus the results in this paper, as ionization is determined mostly by the 
exponential part in Eq. (\ref{eq:ion}) computed with $T_R$, which is a direct 
result of the Monte Carlo simulation.} 
the situation where the electron temperature is 
largely controlled by radiation (Klein \& Castor 1978; Abbott \& Lucy 1985).
Level populations, ionizations and opacities are then updated 
using the temperature structure derived from the Monte Carlo simulation. 
This process is iterated until convergence.

\begin{deluxetable*}{lrccccccc} 
\tablewidth{0pt}
\tablecaption{Properties of the SNe and the spectra}
\tablehead{
Name (group)& 
Epoch \tablenotemark{a} &
$\Delta m_{15}$ \tablenotemark{b} &
$E(B-V)$ \tablenotemark{c} &
$\mu$ \tablenotemark{d} &
$v_{\rm Si}$ \tablenotemark{e} & 
$v_{\rm Ca, ph}$ \tablenotemark{f} & 
$v_{\rm Ca, hv}$ \tablenotemark{g} & 
References 
}
\startdata
SN 2002bo (HVG \tablenotemark{h}) & $ -8.0~$ & 1.17 & 0.38 & 31.67 &15.0 & 14.9 & 22.1 & 1 \\
SN 2002dj (HVG) 		  & $-11.0~$ & 1.12 & 0.10 & 32.98 &16.8 & 17.7 & 27.6 & 2 \\
SN 2002er (HVG) 		  & $ -7.4~$ & 1.33 & 0.36 & 32.90 &13.0 & 15.6 & 23.1 & 3, 4 \\
SN 2001el (LVG \tablenotemark{i}) & $ -9.0~$ & 1.15 & 0.22 & 31.26 &14.0 & 17.1 & 23.8 & 5 \\
SN 2003cg (LVG) 		  & $ -7.6~$ & 1.25 & 1.36 & 31.28 &12.0 & 12.7 & 22.0 & 6 \\
SN 2003du (LVG) 		  & $-11.0~$ & 1.02 & 0.01 & 32.79 &12.0 & 15.5 & 22.5 & 7 \\ 
\enddata
\tablenotetext{a}{Days since B maximum} 
\tablenotetext{b}{Difference of B magnitude between maximum and 15 days later}
\tablenotetext{c}{Color excess 
[sum of that caused by our Galaxy and the SN host galaxy;
$R_V = 3.1$ is adopted except for the host galaxy of SN 2003cg 
($R_V = 1.8, E(B-V) = 1.33 $; Elias-Rosa et al. 2006)]}  
\tablenotetext{d}{Distance modulus (see literature)}
\tablenotetext{e}{Doppler velocity of \ion{Si}{ii} $\lambda$6355 absorption minimum ($10^3$ \kms)} 
\tablenotetext{f}{Doppler velocity of the photospheric component of Ca IR triplet ($10^3$ \kms). see Mazzali et al. (2005b)} 
\tablenotetext{g}{Doppler velocity of the high-velocity component of Ca IR triplet ($10^3$ \kms). see Mazzali et al. (2005b)} 
\tablenotetext{h}{High Velocity Gradient (a group of SNe Ia 
whose Si line velocity declines rapidly 
(see \S \ref{sec:intro}; B05), shown in blue throughout the paper)}
\tablenotetext{i}{{L}ow Velocity Gradient (a group of SNe Ia 
whose Si line velocity declines slowly 
(see \S \ref{sec:intro}; B05), shown in green throughout the paper)}
\tablerefs{(1) Benetti et al. (2004);
(2) Pignata et al. 2005; (3) Pignata et al. (2004); 
(4) Kotak et al. (2005); (5) Mattila et al. (2005);
(6) Elias-Rosa et al. (2006); (7) Stanishev et al. (2007)}
\label{tab:prop}
\end{deluxetable*}

We use the W7 deflagration model (Nomoto et al. 1984) 
as a standard density profile.
Alternatively, a density structure of a delayed detonation model is used 
(model WDD2 in Iwamoto et al. 1999, hereafter, DD model).
The density structures of these models are most different in 
layers above $v = 20,000$ \kms.
The density in the DD model at $v \sim 25,000$ \kms\ is higher than that of 
W7 by a factor of $\sim 10$ (see Fig. 11 in Baron et al. 2006).
Thus the ``density enhancement'' referred to in Mazzali et al. (2005a) is 
equivalent to the structure of the DD model if the densities are increased 
by a factor of $\lsim 10$.
The kinetic energy of the DD model ($1.4 \times 10^{51}$ ergs) is
slightly larger than that of W7 ($1.3 \times 10^{51}$ ergs).

Element abundance distributions (\ie mass fractions of elements) 
are parameterized for better reproduction of the
observed spectra. Initially, a homogeneous abundance distribution is assumed,
averaging the abundance distribution of W7 above the photospheric velocity, and
a general fit of the spectrum is obtained adjusting the abundances. In this
process, the input parameters $L$ and $\vph$ are also optimized. Then a
stratified abundance distribution is adopted and optimized to reproduce the
details. In this paper, we employ two different-abundance zones except for SN
2003cg (3 zones are used; see \S \ref{sec:03cgSi}).

The abundances in the outermost ejecta are further checked against the spectra 
at maximum.
Although the photosphere recedes with time as the ejecta expand, 
the abundances at high velocities can still 
affect the synthetic spectra at later epochs (Stehle et al. 2005).
Therefore we modelled the spectra at maximum using the abundance ratios
derived from the pre-maximum spectra for the region at 
$v > \vph ({\rm early})$, where $\vph ({\rm early})$ is the 
photospheric velocity of the pre-maximum spectra.
If the result is inconsistent (\eg incorrect high-velocity absorptions
resulting from the pre-maximum spectra photospheric component),
the pre-maximum spectra are re-investigated until consistence is achieved.

\section{RESULTS}
\label{sec:results}

\begin{deluxetable}{lrccc} 
\tablewidth{0pt}
\tablecaption{Parameters of the synthetic spectra}
\tablehead{
Name (group) & 
Epoch & 
log($L$) \tablenotemark{a} & 
$\vph$ \tablenotemark{b} &
$T_{\rm ph}$ \tablenotemark{c} 
}
\startdata
SN 2002bo (HVG) & $ -8.0$  & 42.78 & 12,900 & 10,900 \\
SN 2002dj (HVG) & $-11.0$  & 42.44 & 14,000 & 10,200 \\
SN 2002er (HVG) & $ -7.4$  & 42.98 &  9,500 & 17,300 \\
SN 2001el (LVG) & $ -9.0$  & 42.87 & 10,500 & 17,100 \\
SN 2003cg (LVG) & $ -7.6$  & 42.84 &  9,000 & 17,000 \\
SN 2003du (LVG) & $-11.0$  & 42.71 & 10,000 & 18,600 \\ 
\enddata
\tablenotetext{a}{Bolometric luminosity (\ergs) in log} 
\tablenotetext{b}{Photospheric velocity (\kms)} 
\tablenotetext{c}{Photospheric temperature (K) computed by the code 
taking the back-scattering effect into account}
\label{tab:param}
\end{deluxetable}

\begin{deluxetable*}{lrcccccccc} 
\tablewidth{0pt}
\tablecaption{Mass fractions used for the synthetic spectra}
\tablehead{
Name (group) &
Epoch & 
$X$(C)\tablenotemark{a} &
$X$(Si)\tablenotemark{b} &
$X$(S)\tablenotemark{c} &
$X$(Ca)\tablenotemark{d} &
$X_{\rm HV}^{\rm W7}$(Ca)\tablenotemark{e} &
$X_{\rm HV}^{\rm DD}$(Ca) \tablenotemark{f} &
$X$(Fe)$_0$ \tablenotemark{g} &
$X$(\Nifs)$_0$ \tablenotemark{h} 
}
\startdata
SN 2002bo (HVG)  & $-8.0$  & --- & 0.51 & 0.10 & 0.02 &
0.20/0.02 & 0.02/0.006 & 0.020  & 0.045\\

SN 2002dj (HVG) & $-11.0$ & --- & 0.50 & 0.09 & 0.02 &
$>$0.80/0.08 & 0.02/0.005 & 0.005 & 0.0052\\

SN 2002er (HVG) & $-7.4$  &  --- & 0.76  & 0.12 &  0.05 &
0.40/0.05 & 0.05/0.007 & 0.012-0.013  & $<$ 0.025\\

SN 2001el (LVG) & $-9.0$  & 0.003 & 0.42 & 0.05 & 0.02 &
$>$ 0.80/0.10 & 0.10/0.02 & 0.048 & 0.19 \\

SN 2003cg (LVG) & $-7.6$  & $<$ 0.002 & 0.40 (0.08\tablenotemark{i}) & 0.080 & 0.02 & 
0.03/0.005 & 0.001/0.0001 & 0.0033-0.0036 & $<$ 0.0088 \\

SN 2003du (LVG) & $-11.0$ & 0.002  & 0.29  & 0.060 & 0.016 &
0.70/0.11 & 0.10/0.015 &  0.003 & 0.005  \\
\enddata
\tablenotetext{a}{Mass fraction of C at the photosphere} 
\tablenotetext{b}{Mass fraction of Si at the photosphere} 
\tablenotetext{c}{Mass fraction of S at the photosphere} 
\tablenotetext{d}{Mass fraction of Ca at the photosphere} 
\tablenotetext{e}{Mass fraction of Ca in the high-velocity layers 
with the density structure of W7 model without/with CSM interaction} 
\tablenotetext{f}{Mass fraction of Ca in the high-velocity layers
with the density structure of the DD model without/with CSM interaction}
\tablenotetext{g}{Mass fraction of {\it stable} Fe 
(Fe except for \Fefs\ from \Cofs\ decay) at the photosphere}
\tablenotetext{h}{Mass fraction of \Nifs\ at the photosphere 
(at the time of the explosion)}
\tablenotetext{i}{Mass fraction of Si in the intermediate shell at 
$v=17,000 - 20,500$ \kms}
\label{tab:abun}
\end{deluxetable*}

In this section, we present the results of the modeling for each SN.
The properties of the SNe and the spectra are summarized 
in Table \ref{tab:prop}.
In addition to the parameters $L$ and $v_{\rm ph}$, and the abundances 
in the ejecta, the distance modulus ($\mu$) 
and the reddening [in terms of $E(B-V)$ 
and $R_V$] are also required to reproduce the observed spectra. 
Thus the luminosity and the epoch, which are often neglected by rescaling 
the flux arbitrarily, are important because $\vph$ is constrained not only by 
the line velocities but also by the relation 
$L \propto v_{\rm ph}^2 t^2 T_{eff}^4$.
Here $t$ is the time since the explosion and $T_{eff}$ the effective 
temperature. 

All spectra have been calibrated against the photometry. The epoch is estimated
assuming a rise time of 19 days for all objects for simplicity. 
We take $\mu$ and $E(B-V)$ from the literature (see Table \ref{tab:prop}).
Using all the quantities above, 
the photospheric temperature $T_{\rm ph}$ is computed in the code considering 
the back-scattering effect, and thus tends to be higher than 
$T_{eff}$ (Mazzali \& Lucy 1993; Mazzali 2000).

In order to facilitate the comparison of intrinsic colors, all spectra are
displayed after correcting for reddening.  The values of $L$ and $\vph$ derived
from modeling are shown in Table \ref{tab:param} along with the computed
photospheric temperature ($T_{\rm ph}$), while the abundances of the elements
at the photosphere that most affect the synthetic spectra 
are shown in Table \ref{tab:abun}.
The abundances in the outer shell are not strongly constrained
because of the lack of strong lines.
Only \ion{Ca}{ii} has high velocity lines (see following sections), and
the Ca mass fraction in the outer shell is shown in the table ($X_{\rm HV}$(Ca)).
The iron mass fraction $X$(Fe) indicates the sum of stable Fe and \Fefs\
produced by \Nifs\ decay at the time of the observation. If only stable Fe is
considered, this is explicitely denoted as $X$(Fe)$_0$, where the index ''0``
means ''at explosion``. Furthermore, we use $X$(\Nifs)$_0$ for the initial 
mass fraction of \Nifs\ at the time of the explosion.

\subsection{SN 2002bo (Day $-8$, HVG)}

\begin{figure}
\begin{center}
\includegraphics[scale=0.85]{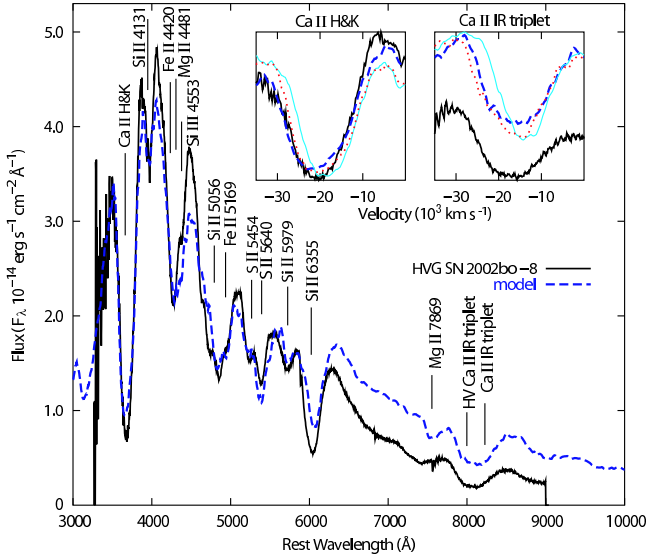} 
\figcaption{
Observed spectrum of the HVG SN 2002bo at day $-8$ (black solid) 
and a model with W7-CSM density/abundance structure (blue dashed).
The left inset shows the \ion{Ca}{ii} H\&K feature (mean $\lambda\sim3950$),
the right inset the \ion{Ca}{ii} IR feature (mean $\lambda\sim8567$). 
The W7-CSM spectrum is compared to W7/homogeneous abundance (thin cyan) and DD
(dotted red) models, both without CSM interaction. In both insets, the 
wavelength scale is replaced by the corresponding Doppler velocies.
\label{fig:02bo}}
\end{center}
\end{figure}

\subsubsection{General Properties}

The well-studied SN 2002bo (Stehle et al. 2005) is a member of the HVG group.
The spectrum at $-8$ days is shown in Figure \ref{fig:02bo}, together with our 
synthetic spectrum. Good agreement is obtained with $\log L$ (\ergs) $ = 42.78$ 
and $\vph = 12,900$ \kms, giving a $\Tph = 10,900$\,K.
The flux redward of $\lambda \sim 6000$ \AA\ is overestimated 
(see also Stehle et al. 2005).
This may be due to a deviation from the blackbody approximation. 
A modified input radiation field, where the blue region ($\lambda \lsim 5000$
\AA) has greater weight, can reproduce the spectrum better, with only a small 
difference in the photospheric temperature and ionization states compared to 
the original case.

\subsubsection{Iron-group Elements}
\label{sec:02boFe}

An absorption feature around 4300 \AA\ results mainly from
\ion{Mg}{ii} $\lambda$4481 and \ion{Fe}{iii} $\lambda$4420, with a smaller
contribution by \ion{Si}{iii} $\lambda$4553.
Around 4800 \AA, many \ion{Fe}{ii} and \ion{Si}{ii} lines, 
\eg \ion{Fe}{ii} $\lambda$5169 and \ion{Si}{ii} $\lambda$5056, 
produce the strong absorption feature.
To account for these strong features, the mass fraction of Fe, 
including both stable Fe and decay products of \Nifs, 
should be $X$(Fe)=0.022 at the photosphere.

\begin{figure}
\begin{center}
\includegraphics[scale=0.85]{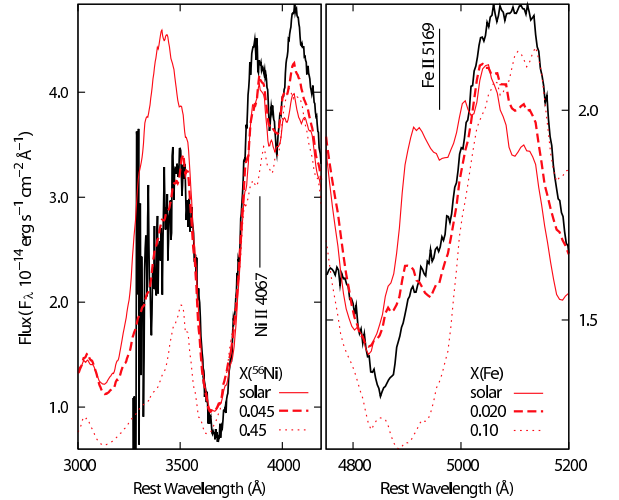} 
\figcaption{
SN 2002bo: Observed spectrum around 3500 \AA\ ({\it left}) and 4900 \AA\ 
({\it right}) compared with synthetic profiles for different metal 
abundances.
\label{fig:02boFe}}
\end{center}
\end{figure}

Our method of estimating the Fe abundance is illustrated in Figure 
\ref{fig:02boFe}, right panel. For solar Fe abundance, the absorption at 
4950 \AA\ (mainly \ion{Fe}{ii} $\lambda$5169) is weaker than the observed 
feature (red solid line).
On the other hand, Fe-rich ejecta [$X$(Fe)$\gsim 0.10$] produce too  
strong a feature (dotted line).

The abundance of radioactive \Nifs\ synthesized in the explosion can be 
estimated using the Ni and Co lines that are present around 2000-4000\AA.
Since at $t \sim$ 10 days around 30\% of \Nifs\ has not yet decayed, 
the strong \ion{Ni}{II} $\lambda$4067 line is useful to constrain 
the mass fraction.
The procedure is illustrated in Figure \ref{fig:02boFe}, left panel.
For a larger \Nifs\ mass fraction [$X$(\Nifs)$_0=0.45$], 
the absorption at 3900 \AA\ 
appears (red dotted line), which is inconsistent with the observed feature.
On the other hand, when only a solar abundance of \Nifs\ is assumed, the 
flux level at $\lambda < 3500$ \AA\ becomes too high (red solid line).
This is because of insufficient blocking of (near) UV photons by iron group 
lines.
[Our code treats the blocking by iron group elements correctly,
considering photon branching (Lucy 1999; Mazzali 2000)].

A \Nifs\ mass fraction $X$(\Nifs)$_0$ = 0.045 is derived.
Although the observed flux level at $\lambda < 3500$ \AA\ is often uncertain, 
the upper limit of $X$(\Nifs)$_0$ is strict because it is derived from 
the absence of \ion{Ni}{II} $\lambda$4067.
Therefore, most Fe detected above the photosphere must be stable 
[$X$(Fe)$_0$= 0.020].

\subsubsection{Silicon and Carbon}

The ratio of the two \ion{Si}{ii} absorptions at 5800\AA\ and 6000\AA\ 
is reproduced well.
A strong \ion{Si}{ii} $\lambda$6355 suggests that the photosphere resides 
in the Si-rich layer.
The mass fraction of Si is estimated as $X$(Si)$\sim$0.50.
The emission peak of \ion{Si}{ii} $\lambda$6355 is well developed,
and there is no evidence of \ion{C}{ii} $\lambda$6578.
The blue wing of the synthetic \ion{Si}{ii} absorption is weak, 
suggesting that the line may be contaminated by a HVF, 
as are the \ion{Ca}{ii} lines (see \S 3.1.4).
At 7500 \AA, a weak absorption is seen, which is often attributed to
\ion{O}{i} $\lambda$7774.
However, in the synthetic spectrum the feature is mainly due to 
\ion{Mg}{ii} $\lambda$7869 and \ion{Si}{ii} $\lambda$7849.
These lines are actually stronger than the \ion{O}{i} line if  
\ion{Mg}{ii} $\lambda$4481 and \ion{Si}{ii} $\lambda$6355 are reproduced.

\subsubsection{Calcium}

The spectrum around \ion{Ca}{ii} H\&K and the \ion{Ca}{ii} IR triplet is 
enlarged in the insets of Figure \ref{fig:02bo}.
The thin cyan line shows a synthetic spectrum computed with the original W7 
density structure and homogeneous abundances, setting $X$(Ca)=0.02.
While the observed features have a blue wing extending to $v \sim 30,000$ \kms\ 
(\ion{Ca}{ii} H\&K) and $27,000$ \kms\ (\ion{Ca}{ii} IR triplet), the wing of 
the synthetic absorption disappears at $v \sim 27,000$ \kms\ and 
$v \sim 23,000$ \kms, respectively.
This difficulty has been reported in many papers, \eg  Hatano et al. (1999) for 
SN 1994D, Gerardy et al. (2004) for SN 2003du, Mazzali et al. (2005a) for 
SN 1999ee and Mazzali et al. (2005b) for the objects in this paper.
To reproduce these features
\footnote{Here the width of the feature and the strength of the absorption 
relative to the continuum are compared because of the poor fit of the 
continuum level.}, 
we test several cases.

First, we modified the abundance distribution, keeping the W7 density structure. 
Although the features can be reproduced with $X$(Ca) $\gsim$ 0.20 at
$v > 22,000$ \kms, such a large fraction is very unlikely because the Ca mass
fraction in the most Ca-rich layer is $\lsim$ 0.1 in all explosion models
(see \eg Iwamoto et al. 1999). 
Secondly, when we use the DD density structure, the
observed profile can be reproduced with $X$(Ca)=0.02 at $v > 22,000$ \kms\
(insets of Fig. \ref{fig:02bo}, dotted red lines) because of the higher
density at $v > 20,000$ \kms.

Finally, we test the interesting possibility that interaction with CSM leads 
to some abundance mixing (Gerardy et al. 2004). 
As discussed by Mazzali et al. (2005a), when a hydrogen-rich CSM is mixed with 
the ejecta, the electron density increases, favouring recombination. 
Therefore some fraction of \ion{Ca}{iii}, which is the dominant state (see 
\S \ref{sec:HVG}), becomes \ion{Ca}{ii}, making the \ion{Ca}{ii} line stronger.
Here we assume that the interaction causes a density jump by a factor of 4 at 
$v > 22,000$ \kms, corresponding to $0.01 \Msun$ of material.
Hydrogen is assumed to be mixed into the shocked region and we find that 
H$\alpha$ does not appear as long as $X$(H) $< 0.30$
($M$(H)$\sim 5 \times 10^{-3} \Msun$)
With this model, the mass fraction of Ca in the high-velocity layers is reduced 
to $X$(Ca)=0.02 (with the W7 density structure; dashed blue line in insets of 
Fig. \ref{fig:02bo}) and $X$(Ca)=0.006 (with the DD density structure), 
respectively. 
The deceleration following the interaction is not considered in our 
computation and a more realistic hydrodynamic calculation is needed to 
confirm this scenario.

\subsection{SN 2002dj (Day $-11$, HVG)}

\begin{figure}
\begin{center}
\includegraphics[scale=0.85]{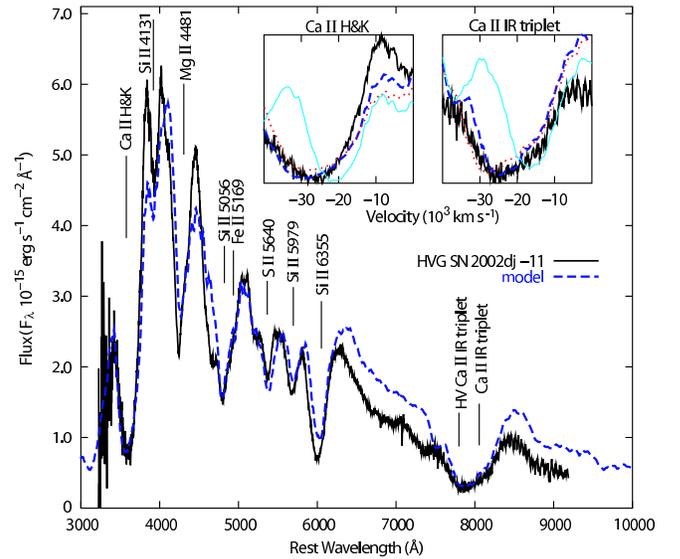} 
\figcaption{
Observed spectrum of the HVG SN 2002dj at day $-11$ (black solid) 
and a model with W7-CSM density/abundance structure (blue dashed).
As in Fig. \ref{fig:02bo}, the insets show \ion{Ca}{ii} H\&K (left) and 
\ion{Ca}{ii} IR (right). The W7-CSM spectrum is compared to W7 (thin cyan) 
and DD (dotted red) models, both with homogeneous abundance and without 
CSM interaction.
\label{fig:02dj}}
\end{center}
\end{figure}

\subsubsection{General Properties}

SN 2002dj has a spectral sequence and LC similar to SN 2002bo 
(Pignata et al. 2005), and is also an HVG SN.
The spectrum $11$ days before B maximum is well reproduced by a synthetic 
spectrum with 
$\log L$ (\ergs) = 42.44 and $\vph = 14,000$ \kms\ (Fig. \ref{fig:02dj}).
Almost all lines are the same as in the spectrum of SN 2002bo, although the 
velocities are higher because the epoch is earlier.
The photospheric temperature, 10,200 K, is also similar.

\subsubsection{Iron-group Elements}

\ion{Fe}{ii} and \ion{Fe}{iii} lines contribute to the features 
near 4200 and 4800 \AA, as in SN 2002bo, but they are weaker.
In fact, the absorption at 4200 \AA\ is caused mostly 
by \ion{Mg}{ii} $\lambda$4481.
Utilizing the feature around 4800 \AA\ (a combination of \ion{Si}{ii} and 
\ion{Fe}{ii} lines) as in \S \ref{sec:02boFe}, the mass fraction 
of Fe is evaluated as $X$(Fe) $\sim 0.005$ at the photosphere.
The features at 3500 and 3900\AA\ require $X$(\Nifs)$_0$ = 0.005.
Therefore, the Fe causing the observed spectral features is again 
mostly stable Fe.
Although the values are smaller than those of SN 2002bo, they are still 
larger than the solar abundances by a factor of 3.5 and 70 for Fe and Ni, 
respectively.

\subsubsection{Silicon and Carbon}

The \ion{Si}{ii} $\lambda$5979 line is quite strong, which is consistent with 
the low photospheric temperature in the synthetic spectrum.
The \ion{Si}{ii} $\lambda$6355 line in SN 2002dj is the broadest of the 6 SNe.
In the model, the mass fraction of Si is $X$(Si)$= 0.50$ at the photosphere. 
The synthetic profile does not show enough absorption in the blue wing, as was 
the case for SN 2002bo, suggesting the presence of a \ion{Si}{ii} HVF.
The emission profile of \ion{Si}{ii} $\lambda$6355 is well developed and there 
is no evidence of an absorption by \ion{C}{ii} $\lambda$6578.

\subsubsection{Calcium}

The \ion{Ca}{ii} HVF in SN 2002dj is the strongest of the 6 SNe. Even assuming
$X$(Ca)=0.80 at the corresponding velocity range, the profile is not reproduced
with the W7 density structure (the thin cyan line in insets of Fig. \ref{fig:02dj} 
is a homogeneous abundance model with $X$(Ca)=0.02). 
The blue dashed line is a synthetic spectrum obtained with a density jump at 
$v > 23,000$ \kms\ (adding 0.005 $\Msun$), where $X$(H) = 0.30 and 
$X$(Ca) = 0.10. This hydrogen mass fraction is an upper limit for the absence of
H$\alpha$ again ($M$(H) $\sim 2 \times 10^{-3} \Msun$). 
Using the DD density structure, the Ca mass fraction at high-velocity layers 
is reduced to $X$(Ca)=0.02 (assuming no CSM interaction, red dotted lines in 
the insets of Fig. \ref{fig:02dj}) and $X$(Ca)=0.005 (assuming CSM 
interaction), respectively.

\subsection{SN 2002er (Day $-7.4$, HVG)}

\begin{figure}
\begin{center}
\includegraphics[scale=0.85]{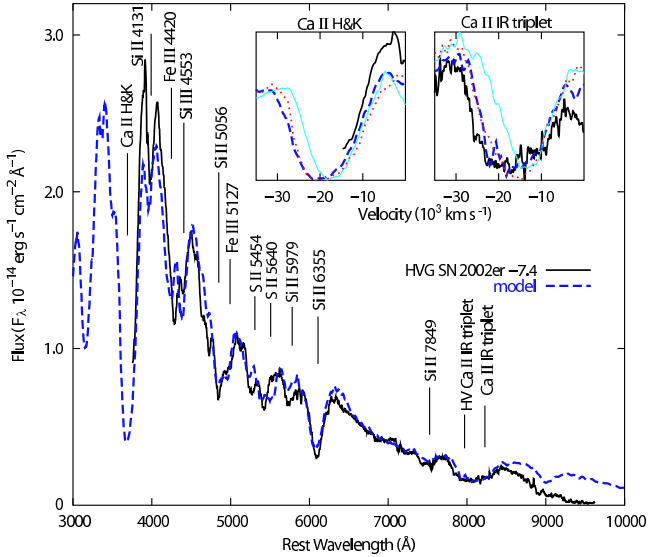}
\figcaption{
Observed spectrum of the HVG SN 2002er at day $-7.4$ (black solid) 
and a model with W7-CSM density/abundance structure (blue dashed).
As in fig. \ref{fig:02bo}, the left inset shows \ion{Ca}{ii} H\&K, the right 
inset \ion{Ca}{ii} IR. The W7-CSM spectrum is compared to W7 (thin cyan) and 
DD (dotted red) models, both with homogeneous abundance and without CSM 
interaction.
\label{fig:02er}}
\end{center}
\end{figure}

\subsubsection{General Properties}

The spectra of the HVG SN 2002er were studied by Kotak et al. (2005). We use 
the earliest spectrum that covers the \ion{Ca}{ii} IR triplet ($-7.4$ days).
A synthetic spectrum with $\log L$ (\ergs) $=  42.98$ and $\vph = 9,500$ \kms\
provides the best fit (Fig. \ref{fig:02er}).
The lower $\vph$ agrees with the fact that this SN has lower line velocities 
than other HVG SNe (Kotak et al. 2005; B05). At the same time, 
$T_{\rm ph}= 17,300$ K, higher than in both SNe 2002bo and 2002dj.
This results from the combination of the higher $L$ and lower $\vph$.

\subsubsection{Iron-group Elements}

\ion{Fe}{iii} lines contribute to the blue absorption near 
4300\AA\ (\ion{Fe}{iii} $\lambda$4396, 4420, 4431).
At 4800\AA, \ion{Fe}{iii} lines (\eg $\lambda$5127) dominate over
\ion{Fe}{ii} lines (\eg $\lambda$5018), 
because the temperature is higher than in SNe 2002bo and 2002dj.
To get sufficiently strong Fe lines, $X$(Fe)=0.013 near the photosphere.
Although the lack of spectral coverage at $\lambda < 3800$\AA\ makes it 
difficult to estimate the abundances of Ni and Co (see \S \ref{sec:02boFe}), 
an upper limit for these elements can be estimated as $X$(\Nifs)$_0<$0.025
considering the absence of \ion{Ni}{ii} $\lambda$4067.
Therefore, the mass fraction of stable Fe is $X$(Fe)$_0=$0.012-0.013.

\subsubsection{Silicon and Carbon}

Around 4300\AA, \ion{Si}{iii} ($\lambda$4553, 4658) lines are stronger than 
in other HVG SNe, indicating a higher temperature.
The blue wing of \ion{Si}{ii} $\lambda$6355 is also reproduced fairly well, 
and there is no clear evidence of a \ion{Si}{ii} HVF, 
unlike SNe 2002bo and 2002dj.
\ion{C}{ii} $\lambda$6578 is not seen, but the \ion{Si}{ii} emission peak 
is weaker than in SNe 2002bo and 2002dj.
A weak absorption near 7500\AA\ is mostly
due to \ion{Si}{ii} $\lambda$7849, and the contribution of 
\ion{Mg}{ii} $\lambda$7869 is small.

\subsubsection{Calcium}

The \ion{Ca}{ii} IR triplet is again not accounted for by the density 
structure of W7 (thin cyan lines in the insets of Fig. \ref{fig:02er} show 
the profile of one-zone abundance model with $X$(Ca)=0.05).
With the original W7 density structure, a Ca-dominated layer 
[$X$(Ca) $\sim$ 0.40] is required for the feature, which is highly unrealistic.
The line depth and width can be reproduced with a density jump at 
$v > 21,000$ \kms, increasing the mass by 0.015 $\Msun$ and 
containing $X$(H) = 0.30 ($M$(H)$\sim 5 \times 10^{-3} \Msun$) 
and $X$(Ca) = 0.05, which is similar to SNe 2002bo and 2002dj.
If the DD model density profile is used instead, the feature is reproduced with 
$X$(Ca)=0.05 at $v > 21,000$ \kms\ even without CSM interaction 
(insets of Fig. \ref{fig:02er}, dotted red lines).
If CSM interaction is considered, this is further reduced to $X$(Ca)=0.007.

\subsection{SN 2001el (Day $-9.0$, LVG)}

\begin{figure}
\begin{center}
\includegraphics[scale=0.85]{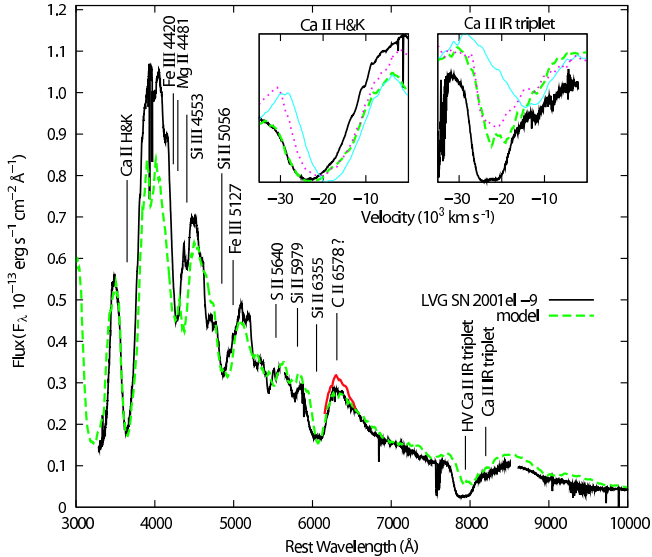}
\figcaption{
Observed spectrum of the LVG SN 2001el at day $-9$ (black solid) and a model 
with W7-CSM density/abundance structure (green dashed). The red solid line 
shows the profile without C in the ejecta.
As in fig. \ref{fig:02bo}, the left inset shows \ion{Ca}{ii} H\&K, the right 
inset \ion{Ca}{ii} IR. The W7-CSM spectrum is compared to W7 (thin cyan) 
and DD (dotted red) models, both with homogeneous abundance and without 
CSM interaction.
\label{fig:01el}}
\end{center}
\end{figure}

\subsubsection{General Properties}

The LVG SN 2001el was studied in spectroscopy (Mattila et al. 2005) and 
spectropolarimetry (Wang et al. 2003).
The best-fitting synthetic spectrum has $\log L$ (\ergs) $= 42.87$ and 
$\vph = 10,500$ \kms\ (Fig. \ref{fig:01el}). 
This leads to $T_{\rm ph} = 17,400$ K, which is similar to SN 2002er, as can 
be expected since the spectra are similar.

\subsubsection{Iron-group Elements}

As in SN 2002er, the features around 4300 \AA\ and 4800 \AA\ are
attributed mostly to \ion{Fe}{iii} lines. 
An iron mass fraction $X$(Fe) $=$ 0.055 at the photosphere is needed 
to reproduce these features.
Relatively reliable flux data at $\lambda \sim 3300-3500$ \AA\ 
makes it possible to estimate the mass fractions of Ni and Co reliably 
(see \ref{sec:02boFe}):
$X$(\Nifs)$_0$=0.19 provides the best fit.
Thus the mass fraction of stable Fe is estimated as $X$(Fe)$_0$=0.048.

\subsubsection{Silicon and Carbon}

The narrow \ion{Si}{iii} feature near 4400 \AA\ has similar strength as in 
SN 2002er, supporting the similar photospheric velocity and temperature.
The \ion{Si}{ii} $\lambda$6355 absorption has a boxy profile and is not 
reproduced well by the synthetic spectrum, suggesting a contribution from HVFs.
Similar profiles were also seen in SNe 1990N 
(Leibundgut et al. 1991; Mazzali 2001) and 2005cf (Garavini et al. 2007).

The \ion{Si}{ii} $\lambda$6355 emission peak is not as round and 
well developed in SN 2001el as in SNe 2002bo and 2002dj.
The synthetic spectrum computed including C 
(Fig. \ref{fig:01el}, dashed green line) reproduces the feature well. 
If C is not included, the emission peak becomes too strong
(Fig. \ref{fig:01el}, red line).
This may be a sign of the presence of C. 
The C abundance required for the feature is, however, only $X$(C) $\sim 0.005$.
The mass of C is estimated as M(C) $\sim 0.02 \Msun$ assuming the W7 density 
structure and a homogeneous C mass fraction above the photosphere.
Since \ion{C}{ii} $\lambda$6578 is the strongest optical C line, the presence 
of this amount of C does not affect other parts of the spectrum.

\subsubsection{Calcium}

As in other SNe, the \ion{Ca}{ii} HVFs cannot be reproduced with the W7 density 
structure (the thin cyan line in insets of Fig.\ref{fig:01el} shows the profile 
of a homogeneous abundance model with $X$(Ca)=0.02).
The HVFs in SN 2001el are very strong and are not reproduced, even using 
$X$(Ca)=0.80 at high velocities.
A better match is obtained using a density jump of a factor of 4 at 
$v \gsim 21,500$ \kms, adding a mass of 0.015 $\Msun$, mixed-in hydrogen 
($X$(H)=0.30, $M$(H)$\sim 5 \times 10^{-3} \Msun$), and $X$(Ca)=0.10 (insets 
of Fig \ref{fig:01el}, dashed green line), but the synthetic \ion{Ca}{ii} IR 
triplet is too narrow.
It may be made broader by increasing Ca mass fraction 
or using a larger density jump, but both 
changes seem unlikely and lead to a worse fit of \ion{Ca}{ii} H\&K.
A large Ca mass fraction, $X$(Ca)=0.10 in the high-velocity layers, 
is needed even with the DD density structure (insets of Fig. \ref{fig:01el}, dotted red lines).
If CSM interaction is considered for the DD model, this is, however, 
reduced to $X$(Ca)=0.02.

\subsection{SN 2003cg (Day $-7.6$, LVG)}

\begin{figure}
\begin{center}
\includegraphics[scale=0.85]{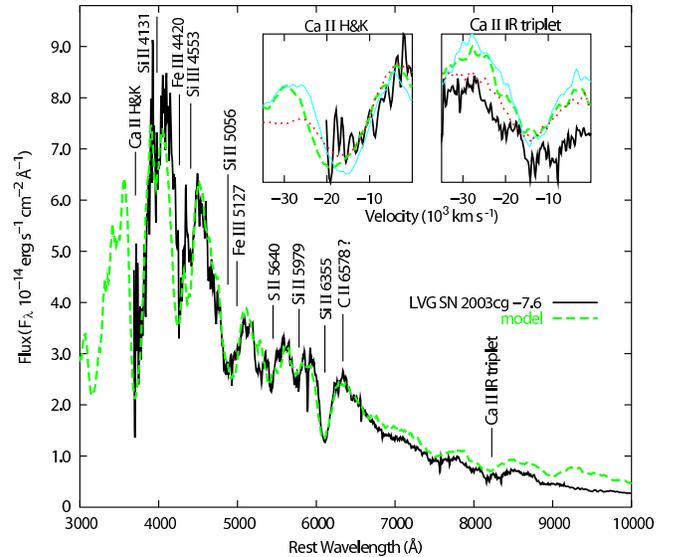}
\figcaption{
Observed spectrum of the LVG SN 2003cg at day $-7.6$ (black solid) and a model 
with W7-CSM density/abundance structure (green dashed). This model uses  
stratification with three abundance zones (\S \ref{sec:03cgSi}).
As in Fig. \ref{fig:02bo}, the left inset shows \ion{Ca}{ii} H\&K, the right 
inset \ion{Ca}{ii} IR. The W7-CSM spectrum is compared to W7 (thin cyan) 
and DD (dotted red) models, both with homogeneous abundance and without 
CSM interaction.
\label{fig:03cg}}
\end{center}
\end{figure}

\subsubsection{General Properties}
The LVG SN 2003cg was studied by Elias-Rosa et al. (2006).
A synthetic spectrum with $\log L$ (\ergs) $=42.84$ and $\vph=9,000$\,\kms\ 
provides the best fit (Fig. \ref{fig:03cg}).
The model has $\Tph=17,000$\,K, similar to SNe 2002er and 2001el  
but higher than SNe 2002bo and 2002dj.

\subsubsection{Iron-group Elements}

To account for the strength of the \ion{Fe}{ii} and \ion{Fe}{iii} 
features at 4000-4900 \AA, $X$(Fe)$\sim$ 0.036.
We estimate an upper limit for \Nifs\ of $X$(\Nifs)$_0< 0.066$
from the absence of the \ion{Ni}{ii} $\lambda$4067 line.
The mass fraction of stable Fe is therefore $X$(Fe)$_0= 0.033-0.036$. 
A similar result is obtained in Elias-Rosa et al. (2006), but
the mass fraction of stable Fe they find is only $X$(Fe)$_0$=0.015.
The main difference is that $\vph$ is lower in our model,
leading to a higher temperature and suppressing \ion{Fe}{ii}. 
In our model, the fraction of \ion{Fe}{ii} is smaller than 
in Elias-Rosa et al. (2006) by a factor of 3.

\subsubsection{Silicon and Carbon}
\label{sec:03cgSi}

SN 2003cg has the narrowest \ion{Si}{ii} $\lambda$6355 line of our 6 SNe, 
although the line velocities are typical for LVG (Elias-Rosa et al. 2006).
To reproduce the narrow feature, we use a layer at 
$v \sim 17,000 - 20,500$ \kms\ with a smaller Si mass fraction 
[$X$(Si) $= 0.08$] than at the photosphere [$X$(Si) $= 0.40$].
There is no evidence of a \ion{Si}{ii}  HVF, which seems consistent with the 
weakness of the \ion{Ca}{ii} HVF (\S 3.5.4).
The emission peak is somewhat suppressed, as in SNe 2002er and 2001el.
Elias-Rosa et al. (2006) identified this as \ion{C}{ii} $\lambda$6578, who   
obtained a strong synthetic \ion{C}{ii} line useing $X$(C)=0.05.
We find that $0.002$ is an upper limit at the photosphere.
However, C is not necessarily needed and the identification is not conclusive.

\subsubsection{Calcium}

While the \ion{Ca}{ii} IR HVF might be present in SN 2003cg, it is the 
weakest of the 6 SNe.
In fact, a synthetic spectrum computed with the W7 density structure and 
homogeneous abundances [$X$(Ca)=0.02] (insets of Figure \ref{fig:03cg}, 
thin cyan line) provides a reasonably good match to the observed profile,
although a better match is obtained for a slightly higher Ca abundance 
[$X$(Ca)=0.03] at high velocity than at the photosphere [$X$(Ca)=0.02].
In the context of CSM interaction, \ie a density jump by a factor of 4 
and $X$(H)=0.30 ($M$(H)$\sim 5 \times 10^{-3} \Msun$) at $v \gsim 21,000$ \kms, 
$X$(Ca)=0.001 is enough for the HVF.
With the DD density structure a smaller fraction of Ca is required, 
as expected [$X$(Ca)=0.001 (0.0001) without (with) CSM interaction].

\subsection{SN 2003du (Day $-11.0$, LVG)}

\begin{figure}
\begin{center}
\includegraphics[scale=0.85]{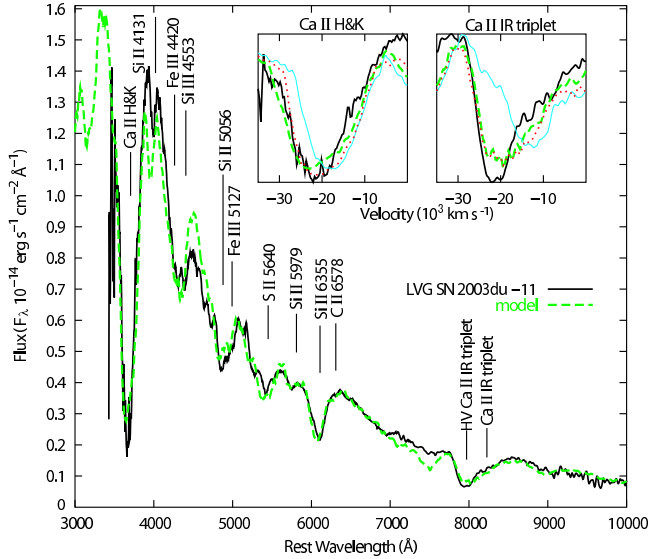}
\figcaption{
Observed spectrum of the LVG SN 2003du at day $-11$ (black solid) and a model 
with W7-CSM density/abundance structure (green dashed).
As in Fig. \ref{fig:02bo}, the left inset shows \ion{Ca}{ii} H\&K, the right 
inset \ion{Ca}{ii} IR. The W7-CSM spectrum is compared to W7 (thin cyan) and 
DD (dotted red) models, both with homogeneous abundance and without 
CSM interaction.
\label{fig:03du}}
\end{center}
\end{figure}

\subsubsection{General Properties}

The LVG SN 2003du was studied by Stanishev et al. (2007).
A synthetic spectrum with $\log L$ (\ergs) $=42.71$ and $\vph = 10,000$ \kms\ 
gives the best agreement with the observed spectrum (Fig. \ref{fig:03du}).
These parameters lead to $T_{\rm ph}=18,600$ K, which is close to that of 
other LVG SNe.

\subsubsection{Iron-group Elements}

The \ion{Fe}{iii} feature near 4300 \AA\ is weaker than in SNe 2002el and 
2003cg, but the temperature is not significantly different, suggesting a 
smaller Fe abundance at the photosphere: $X$(Fe) = 0.003 (determined as in 
\S \ref{sec:02boFe}) is enough to obtain this feature.
We estimate the Ni mass fraction as $X$(\Nifs)$_0$=0.005 from the feature at 
3900 \AA.
Thus, almost all the Fe contributing to the spectrum is likely to be stable.

\subsubsection{Silicon and Carbon}

The \ion{Si}{ii} $\lambda$6355 line is relatively weak and can be 
reproduced with a Si mass fraction of $X$(Si)=0.3.
In spite of the low velocity of the absorption minimum, the blue wing of the 
absorption extends to $v \sim 20,000$ \kms.
Although this could be a sign of high-velocity material as in SNe 2002bo, 
2002dj and 2001el, the line profile is different from other SNe, and the 
high-velocity absorption ($v \sim 15,000 - 20,000$ \kms) is relatively weak
(Mazzali et al. 2005b).

The emission peak of the line is suppressed, as seems common among LVG SNe.
Including carbon with $X$(C) $\sim$ 0.003 improves the fit. If the 
identification is correct, the mass of C is estimated to be M(C)$\sim0.001\Msun$
assuming the W7 density structure and a homogeneous C mass fraction
above the photosphere.

\subsubsection{Calcium}

The \ion{Ca}{ii} IR triplet at day $-5$ was studied by Gerardy et al. (2004).
The spectrum at day $-11$ also shows a conspicuous HVF (Fig. \ref{fig:03du}), 
with profile and strength similar to those of SN 2001el. A very large 
high-velocity Ca mass fraction [$X$(Ca)$>$0.7] is required if W7 is used 
(thin cyan lines in the insets of Fig. \ref{fig:03du} show the profile of a 
homogeneous abundance model with $X$(Ca)=0.016).
For the CSM interaction scenario, we need $X$(Ca)=0.11 at high velocies 
($v \gsim 22,000$ \kms), where the density is increased by a factor of 4, 
adding 0.012 $\Msun$, and $X$(H) is assumed to be 0.30 
($M$(H)$\sim 5 \times 10^{-3} \Msun$, dashed green line in 
insets of Fig. \ref{fig:03du}).
Even if the DD density structure is adopted, a high mass fraction of Ca at high 
velocity is required [$X$(Ca)=0.10]. This is reduced to $X$(Ca)=0.015 when CSM 
interaction is assumed(see Table \ref{tab:abun}).

\section{DISCUSSION}
\label{sec:discussion}

\begin{figure}
\begin{center}
\includegraphics[scale=0.85]{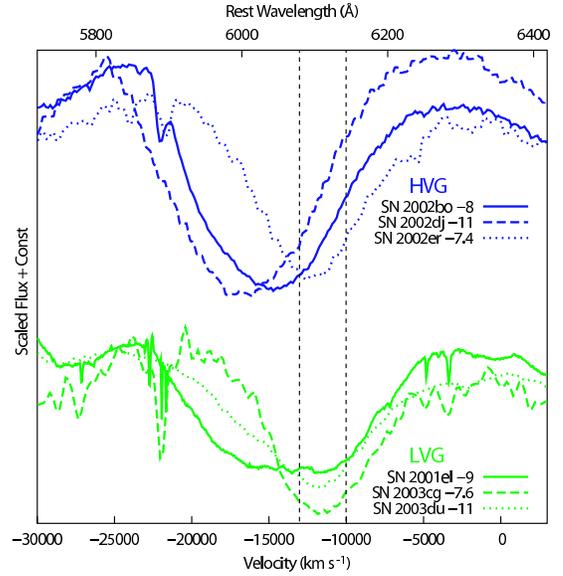}
\figcaption{
Observed \ion{Si}{ii} $\lambda$6355 features of 3 HVG SNe Ia (top, blue) 
and 3 LVG SNe Ia (bottom, green).
Velocities and shapes of the blue wing differ among the objects. 
They range from $15,000$ \kms\ to $25,000$ \kms\ irrespective of the SN group 
(HVG, \S \ref{sec:HVG} or LVG, \S \ref{sec:LVG}).
The black dashed lines correspond to Doppler velocities of $v = 10,000$ and 
$13,000$ \kms, which are the typical photospheric velocities of LVG and HVG, 
respectively.
\label{fig:si}}
\end{center}
\end{figure}

\begin{figure*}
\begin{center}
\includegraphics[scale=0.95]{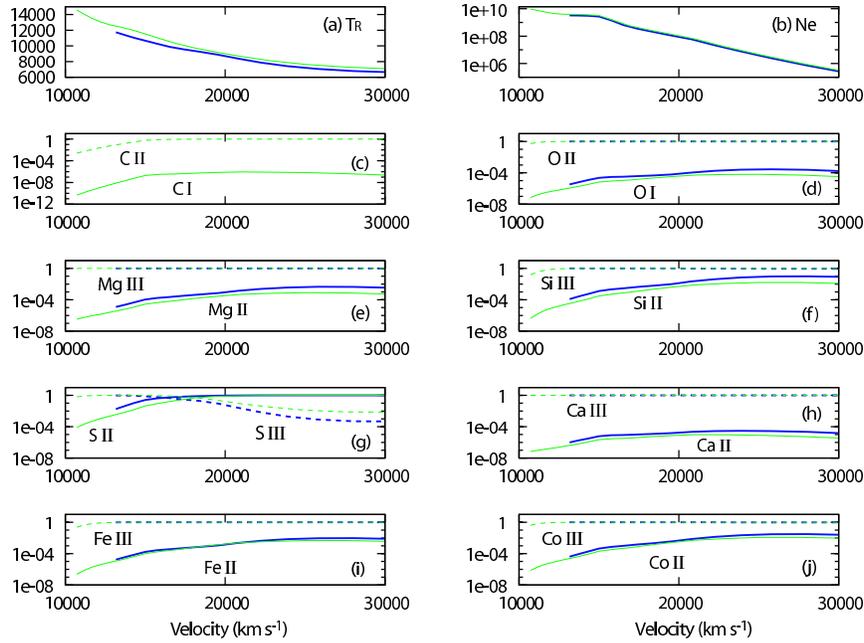}
\figcaption{
Radial distribution of (a) radiation temperature (K),
(b) number density of electrons (${\rm cm^{-3}}$) 
and (c)-(j) ionization fraction of each element.
Blue lines show the values for a HVG SN (2002bo) 
and green lines for a LVG SN (2001el).
The two dominant ionization states are shown inside each panel; solid lines 
represent the lower state (\eg \ion{C}{i}, \ion{O}{i}, \ion{Mg}{ii}).
In panel (c), no blue lines appear, because C is not included in the models of 
HVG SNe.
\label{fig:ion}}
\end{center}
\end{figure*}

We have modelled the pre-maximum spectra of 6 different SNe Ia 
(see Table \ref{tab:prop}).
Three of these (2002bo, 2002dj and 2002er) are classified as High Velocity 
Gradient (HVG), and are characterized by a large Si line velocity that evolves 
rapidly after maximum brightness. 
The other three SNe (2001el, 2003cg, and 2003du), classified as Low Velocity 
Gradient (LVG), have a smaller Si line velocity that evolves slowly after 
maximum.
The parameters of our models are shown in Tables \ref{tab:param} (luminosity 
and photospheric velocity) and \ref{tab:abun} (abundances).

In the following, we summarize the properties of HVG (\S \ref{sec:HVG}) 
and LVG (\S \ref{sec:LVG}) SNe.
Their different appearance stems basically 
from a different photospheric velocity.
The presence of iron-peak elements in the outer layers is revisited
and its implications are discussed in \S \ref{sec:Fe}.
Then, the origin of the difference between HVG and LVG is discussed in 
\S \ref{sec:HVGLVG}.
The variety of High Velocity Features (HVFs) is very large independently of
the classification of HVG or LVG.
The causes for HVFs are also discussed in \S \ref{sec:HVF}.
Finally, possible influences to the LC width-luminosity relation are
discussed (\S \ref{sec:WLR}).

\subsection{The Properties of HVG SNe}
\label{sec:HVG}

HVG SNe 2002bo and 2002dj  have the highest line velocities and the lowest
photospheric temperatures of the 6 SNe.  The higher temperature for SN 2002er
results from the high $L$ and the low $\vph$. SN 2002er seems to be
intrinsically more luminous than SN 2002bo at a similar epoch (Table 2),
although it has a larger decline rate.  This is also supported by the analysis
of nebular spectra (Mazzali et al. 2007).  The velocity of SN 2002er is
intrinsically lower than that of other HVG SNe (B05; Kotak et al. 2005). 
Although a SN may be classified as HVG, its $T_{eff}$ can be as high as that of
LVG SNe if the SN has both a high enough $L$ and low enough $\vph$. This
suggests continuity between two groups (Branch et al. 2006, 2007).

The \ion{Si}{ii} $\lambda$6355 lines of SNe 2002bo and 2002dj have 
well-defined P-Cygni profiles with blue wings reaching $v \sim 23,000$ \kms\
(blue lines in Fig. \ref{fig:si}). In the spectra of these SNe, \ion{C}{ii} is
not seen below at least $v = 20,000$ \kms.  This suggests that the abundance
distribution of HVG SNe resembles that of 1D delayed detonation models
(Khokhlov et al. 1991a; H\"oflich \& Khokhlov 1996; Iwamoto et al. 1999) rather
than that of the 1D deflagration model W7 (Nomoto et al. 1984).

The computed physical conditions in the ejecta of a representative HVG SN,
2002bo, and those of a typical LVG SN of similar $\Delta m_{15}$, and thus 
presumably similar luminosity, SN 2001el, are shown in Figure \ref{fig:ion}. 
In the ionization plot, solid lines represent the lower of two states marked in
each panel. The temperature structure, the number density of electrons, and the
ionization state of SN 2002bo and SN 2001el computed under the assumptions
described in \S \ref{sec:method} are rather similar at all layers with $v >
12,900$\,\kms.  Consequently, the ionization fractions are also similar.

B05 show the evolution of the parameter $\cal R$(Si), the ratio of the 
\ion{Si}{ii} lines at 5800 and 6000 \AA.
$\cal R$(Si) is thought to be a temperature indicator (Nugent et al. 1995).
At pre-maximum epochs ($\lsim -10$ days), $\cal R$(Si) is larger in HVG
than in LVG SNe (see Fig. 2 in B05). 
As we have shown, a typical HVG SN has a lower photospheric temperature before 
maximum.
The fact that SNe 2002bo and 2001el have a similar $\Delta m_{15}$, and 
therefore should have a similar luminosity, suggests that the lower temperature 
of SN 2002bo at pre-maximum epochs reflects its higher photospheric velocity.
Therefore, $\cal R$(Si) is not a good indicator of SN luminosity at  
pre-maximum phases, when the dispersion of the photospheric velocity among 
different SNe is still very large (B05).
On the other hand, since the diversity in photospheric velocity is small at 
epochs near maximum, both $\cal R$(Si) 
and the temperature more closely reflects the intrinsic luminosity.

\subsection{The Properties of LVG SNe}
\label{sec:LVG}

The three LVG SNe have similar photospheric velocities and temperatures.
The temperature structure, electron density, and ionization state of 
SN 2001el are shown in Figure \ref{fig:ion} (green lines).
The run of all physical quantities is similar to HVG SN 2002bo at comparable 
velocities, but the values at the photosphere are significantly different 
because $\vph$ is lower for LVG.
In particular, important ions such as \ion{Si}{ii}, \ion{S}{ii}, \ion{Ca}{ii}, 
and \ion{Fe}{ii} are a factor of $> 10$ less abundant near the photosphere of 
LVG SNe (at $v \sim 10,000$ \kms) then near the photosphere of HVG SNe 
(at $v \sim 13,000$ \kms).

The \ion{C}{ii} $\lambda$6578 line is likely to be present in the LVG SNe
2001el and 2003du, and possibly also in SN 2003cg near the photosphere 
($v \sim 10,000$ \kms). The mass fraction of C required to suppress the 
\ion{Si}{ii} $\lambda$6355 emission peak is however only 0.002-0.003.
The mass of C in SNe 2001el and 2003du is only $\sim 0.001 \Msun$ if we 
assume a homogeneous, spherically symmetric distribution above the photosphere.
Even if we consider that we cannot estimate the C abundance
at $v \gsim 18,000$ \kms\ because of the blend with the \ion{Si}{ii} line, 
and assume $X$(C)=0.5 at $v > 18,000$ \kms, 
the mass of C is at most $\sim 0.01 \Msun$. 
This is still less than predicted by the W7 model ($\sim 0.05 \Msun$).

Although \ion{C}{i} is used to estimate C abundance in Marion et al. (2006),
\ion{C}{i} is not the dominant ionization state as shown in Figure 
\ref{fig:ion} (solid line), and the exact amount of \ion{C}{i} is very
sensitive to the temperature structure. 
In our calculations, \ion{C}{i} lines never become visible even if \ion{C}{ii}
makes a deep absorption, suggesting that \ion{C}{ii} is a better indicator of
carbon abundance in this temperature range.

Alternatively, the emission peak of \ion{Si}{ii} $\lambda$6355 may be
suppressed if Si is detached from the photosphere (see Branch et al. 2002 for 
the case of H$\alpha$). This is however not the case for the objects presented 
here because the velocity of \ion{Si}{ii} decreases to $v < 9,000$ \kms\ at 
subsequent epochs and Si is abundant at the corresponding velocities. 
In addition to this, the fact that the emission profile of the \ion{Si}{ii} 
line is well-peaked at maximum light even in LVG suggests that the suppression 
is related to the conditions of the outer layers.

\subsection{Fe-group elements in the outer ejecta}
\label{sec:Fe}

Both \ion{Fe}{ii} and \ion{Fe}{iii} lines are clearly present in pre-maximum
spectra.  We showed that about a few percent of Fe in mass fraction, including
both stable Fe and decay products of \Nifs, are required to match the observed
lines (\S \ref{sec:02boFe}). Lines of both ions (\ion{Fe}{ii} and
\ion{Fe}{iii}) are correctly reproduced, which suggests that the computation of
the ionization state of Fe is reliable. The fact that Fe is not highly abundant
indicates that the photosphere resides outside the Fe-rich layers.

In order to distinguish stable Fe (\eg $^{54}$Fe) from the decay product of
\Nifs\ (\ie $^{56}$Fe), the abundances of Ni and Co were quantified from the
spectra (\S \ref{sec:02boFe}). In 4 of our SNe, we are able to estimate the
\Nifs\ abundance near the photosphere, while only upper limits could be
obtained in the remaining 2 cases. If \Nifs\ accounted for all the present Fe
abundance at such an early epoch (when only 3\% of \Nifs\ has decayed to
\Fefs), the ejecta would be dominated by \Nifs. This is, however, clearly
inconsistent with the observed spectra and the parameters obtained from the
modeling. Therefore, we conclude that not only \Nifs, but also stable Fe must
exist at the photosphere. The respective abundances obtained are always larger
than solar suggesting that the origin is not the progenitor but nucleosynthesis
during the explosion.

The metal abundances in the outer layers suggest that the abundances are  
not sharply stratified with velocity. 
Even in 1D, the region where \Nifs\ is synthesized extends to Si-rich layers 
because the temperature changes smoothly during the explosion.
$^{54}$Fe can also be synthesized in Si-rich layers (\eg Iwamoto et al. 1999).
Another explanation for Fe-group elements in outer layers may be large 
scale mixing, as seen in recent three-dimensional deflagration models 
(\eg Gamezo et al. 2003; R\"opke \& Hillebrandt 2005; R\"opke et al. 2006). 

The presence of Fe-group elements in the outer layers and the diversity of
their abundances and a low abundance of unburned elements may hinder the use of 
pre-maximum spectra as indicators of the progenitor metallicity (Lentz et al. 
2000).
In fact, larger diversity than predicted by model computations is seen in 
UV spectra (Ellis et al. 2007).
This may reflect diversity in the abundances of Fe-group elements in the outer 
layers, or in the photospheric temperature (\S \ref{sec:HVG} and \ref{sec:LVG}).

\subsection{The Origin of the Difference between HVG and LVG SNe}
\label{sec:HVGLVG}

The difference between HVG and LVG SNe can be studied comparing the properties
of SNe with similar decline rates ($\Delta m_{15}$). For an optimal comparison
at a similar epoch, we again take SNe 2002bo (HVG; $-8.0$ days, $\Delta m_{15}
= 1.17$) and 2001el (LVG; $-9$ days, $\Delta m_{15} = 1.15$) as examples. 

The marginal detection of carbon in the emission peak of 
\ion{Si}{ii} $\lambda$6355 in LVG SNe suggests that
the burning is less powerful in LVG SNe than in HVG SNe. 
It is also interesting that other SNe where C was detected (SN 1998aq, 
Branch et al. 2003; and possibly SN 1994D, Branch et al. 2005) are also LVG
\footnote{Since the number of LVG seems to be larger than that of HVG 
(B05; Mazzali et al. 2007), this could be a statistical effect.}.
In addition, we find that HVG SNe do not have carbon at least 
up to $v\sim 20,000$ \kms.

If more burning occurs in HVG than in LVG SNe, the former may have more 
kinetic energy.
This would tend to make the photospheric velocity of HVG higher. The kinetic 
energy difference between HVG and LVG can estimated by a simple analysis.
We assume that the outermost layer are fully burned to Si in HVG while half 
of the material at $v > 18,000$ \kms\ is oxygen and carbon in LVG.
Since the mass above $v=18,000$ \kms\ is $\sim 0.02 \Msun$ assuming the W7 
density structure, the difference in nuclear energy release is
$\sim 0.02 \times 10^{51}$ erg, which is less than $\sim 2\%$ of the typical 
kinetic energy of SNe Ia models.
The expected difference in photospheric velocity is however only $\lsim 1$\%.
Therefore, it seems unlikely that the difference between HVG and LVG can 
be accounted for by a different kinetic energy alone. 
Yet, such a difference in kinetic energy cannot be ruled out.

Alternatively, HVG SNe may have more massive ejecta than LVG SNe.  
The limiting mass of a spinning WD can exceed the Chandrasekhar mass of the 
static WD, $\sim 1.38 \Msun$ (Uenishi, Nomoto \& Hachisu 2003; 
Yoon \& Langer 2004, 2005; Dom\'inguez et al. 2006).  
Since the density structure of a rotating WD is flatter than that of a static 
one, the photospheric velocity may be larger before maximum but then evolve 
more rapidly.  Unless the WD mass is extremely large -- such as the 2$\Msun$ 
suggested for SN 2003fg by Howell et al. (2006), the expansion kinetic 
energy would not be significantly reduced by a higher binding energy.
However, there is no evidence that HVG SNe tend to synthesize more \Nifs, 
which does not support this interpretation.

Even though the mass and the kinetic energy of the ejecta are similar in HVG 
and LVG, a different Fe-group abundance in the outer region could make their 
appearance different.
Some amount of Fe-group elements is present in the outer layers of all 6
SNe (see \S \ref{sec:Fe}). 
Line velocities are expected to be higher if more Fe-group elements are present
in the outer layers because of the larger opacity of these elements.

Our results show that the near-photospheric abundance of Fe-group elements is
not smaller in HVG than in LVG, even though the photospheric velocity is higher
in HVG by $\sim 2-3,000$ \kms. This might imply that HVG SNe have more Fe-group
elements if the metal abundance is a decreasing function of radius (velocity).
Therefore, the abundances of Fe-group elements in the outer layers can be a
cause of spectral diversity, while differences in the kinetic energy or the
progenitor mass are less likely. To confirm this hypothesis, the metal
abundance in the spectra of LVG with $\vph \gsim 13,000$ \kms\ should be
investigated with earlier spectra.

\subsection{High-Velocity Features}
\label{sec:HVF}

\begin{figure}
\begin{center}
\includegraphics[scale=0.85]{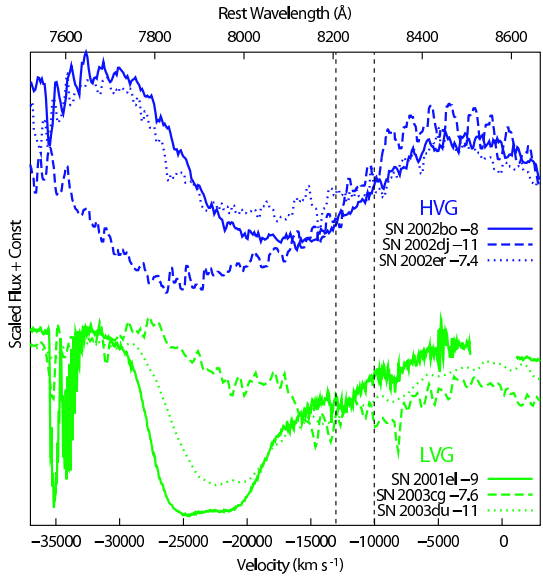}
\figcaption{
\ion{Ca}{ii} IR triplet profiles (a mean wavelength of 8567 \AA\ is assumed) 
of 3 HVG SNe Ia (top, blue) and 3 LVG SNe Ia (bottom, green).
In LVG SNe (SNe 2001el and 2003du in our sample and SN 2003kf in Mazzali et al. 
2005b), HVFs tend to be detached, which is likely the consequence of the high 
photospheric temperature and the low photospheric velocity 
(see \S \ref{sec:HVF}).
On the other hand, the strength of the high-velocity absorption 
seems to be random (Tanaka et al. 2006). 
The dashed black, vertical lines correspond to Doppler velocities of
$v = 10,000$ and $13,000$ \kms, which are the typical photospheric velocities
of LVG and HVG, respectively. 
\label{fig:ca}}
\end{center}
\end{figure}

High-velocity features ($v \gsim 20,000$ \kms) are seen in both \ion{Ca}{ii}
H\&K and in the IR triplet. Except for SN 2003cg, HVFs are not present in 
synthetic spectra based on W7. They can be reproduced using the DD density
structure and a relatively large Ca abundance. Alternatively, they can be also
explained by an enhancement in density of the outer layer ($v > 20-22,000$
\kms) of a factor of 4 from the W7 value assuming some H ($\sim 30$\% in mass
fraction) in this region. Interaction with CSM may produce the latter
situation; mixing of H into the ejecta increases the \ion{Ca}{ii} fraction by a
factor of 5-10 as it favors recombination (Mazzali et al. 2005a).

The profiles of the \ion{Ca}{ii} IR triplet show variations (Fig.
\ref{fig:ca}).  In HVG SNe (blue lines), the profile is round and the 
high-velocity component tends to blend with the photospheric component. In LVG
SNe, on the other hand, the profiles are less blended because of the weakness 
of the photospheric component (green lines, see also SN 2003kf in Mazzali et
al. 2005b). This is a natural consequence of the high $\Tph$ and the low 
$\vph$ of LVG SNe.  The high temperature suppresses \ion{Ca}{ii} and
\ion{Si}{ii} lines near the photosphere, and the low photospheric velocity
makes the position of the photospheric component redder. The combination of
these two effects results in a weaker and redder photospheric absorption, so
that the high-velocity component becomes more detached in LVG SNe.

The strength of the high-velocity component also varies from SN to SN. As our
models show, this is not a temperature effect, because the temperature in the
outer region is similar in all cases (Fig. \ref{fig:ion}).  Although we could
reproduce the strength of the HVF increasing the Ca abundance, the abundances
that are required show a large spread (of a factor of 100; Table
\ref{tab:abun}). This diversity may be a line-of-sight effect if HVFs are
caused by aspherical structures, like a torus or clumps (Tanaka et al. 2006),
as also suggested by the high polarization level of the feature (Wang et al.
2003; Kasen et al. 2003).

The very strong, broad blue wing of the \ion{Si}{ii} $\lambda$6355 absorption 
in SNe 2002bo, 2002dj and 2001el is not perfectly explained by our synthetic 
spectra (Figures \ref{fig:02bo}, \ref{fig:02dj} and \ref{fig:01el}). This  
suggests the presence of HVFs in \ion{Si}{ii} as well as in \ion{Ca}{ii} lines,
although the velocity of the blue edge of the \ion{Si}{ii} absorption is lower
than that of the \ion{Ca}{ii} lines (Figures \ref{fig:si} and \ref{fig:ca}).
On the other hand, SNe 2002er, 2003cg and 2003du lack this feature (Figures
\ref{fig:02er}, \ref{fig:03cg}, and \ref{fig:03du}).

The origin of HVFs is debated (\eg Hatano et al. 1999; Wang et al. 2003; Thomas
et al. 2004; Gerardy et al. 2004; Mazzali 2005a; Kasen \& Plewa 2005; Tanaka et
al. 2006). The difficulty in using the DD model is that it still requires a high
Ca abundance in the high-velocity layers ($v \gsim 23,000$ \kms; Table
\ref{tab:abun}) despite the absence of a very strong Si absorption at similar 
velocities. One problem with the CSM interaction scenario is that it  requires
the accumulation  of a relatively large mass. The estimated mass in the
high-velocity shell is $\sim 0.01 \Msun$. In order to accumulate this mass, a
mass loss rate $\dot{M} \sim 10^{-4}\ \Msun\ {\rm yr}^{-1}$ is required for a
wind velocity of 10 \kms. This mass loss rate is higher than that constrained
from X-ray, optical and radio observations (\eg Cumming et al. 1996; Mattila et
al. 2005; Immler et al. 2006;  Hughes et al. 2007; Panagia et al. 2006). A
combination of these two scenarios  (\ie DD model and CSM interaction) may relax
these requirements (Gerardy et al. 2004).

Alternatively, HVFs may reflect the pre-SN abundances of the progenitor WD.
In the single-degenerate scenario, the WD accretes matter from a companion star
(see Nomoto et al. 1994 for a review).
At an accrretion rate suitable for stable H shell burning (Nomoto 1982),
He shell flashes are rather weak (Taam 1980; Fujimoto \& Sugimoto 1982),
and their products can accrete onto the WD.

The products of He shell flashes could be Ca-rich depending on conditions 
such as the pressure at the burning shell (Hashimoto, Hanawa \& Sugimoto 1983).
After many cycles of He flashes, when the mass of the WD reaches the 
Chandrasekhar mass, the outermost layers of the WD should consist not only
of C and O but also of heavier elements.
If such matter resides in the outermost layers after the explosion,
it could cause HVFs.
To investigate this scenario further, the growth of the WD should be studied 
in detail.

If \ion{Si}{ii} and \ion{Ca}{ii} HVFs are caused by the explosion itself or by 
CSM interaction, they should correlate (see Fig. 12 in Tanaka et al. 2006).
In fact, this seems to be the case for SNe 2002bo, 2002dj, 2001el, and 2003cg, 
but the correlation is not clear for SNe 2002er and 2003du
\footnote{Si and Ca HVF may correlate in an earlier spectrum of SN 2003du. 
See Stanishev et al. (2007).}
(Figs. \ref{fig:si} and \ref{fig:ca}).
This may be useful to discriminate the origin of the HVFs.

\subsection{Implications for the Width-Luminosity Relation}
\label{sec:WLR}

The relation between LC width and luminosity in SNe Ia indicates that the 
properties of SNe Ia are determined by a single dominant parameter.
However, dispersion is present in the relation, and it may due to 
uncertainties in distance and reddening or to intrinsic properties of 
SNe Ia (Mazzali \& Podsiadlowski 2006).
Understanding the origins of the dispersion is important for 
a more precise cosmological use of SNe Ia.

The earliest spectra of SNe Ia are clearly not determined by a single
parameter, \eg SNe with a similar maximum luminosity have different line
velocities. Therefore, the variation in the early phase spectra could, to some
extent, reveal an intrinsic dispersion in the width-luminosity relation. Since
the variation of properties is larger in HVG than in LVG SNe (B05; Mazzali et
al. 2007), the HVG group at least could be related to the dispersion in the
width-luminosity relation.

A first parameter to look at in this context is the kinetic energy of the
ejecta. HVG SNe may have higher kinetic energy than LVG by at most $\sim 2$\% 
(\S \ref{sec:HVGLVG}). Since the light curve width ($\tau_{\rm LC}$) scales
roughly as $\tau_{\rm LC} \propto \KE^{-1/4}$  (Arnett 1982), where $\KE$ is
the kinetic energy of the ejecta, the LC width of HVG SNe may become narrower
than that of LVG SNe. This effect should not be significant 
given the dependence of $\tau_{\rm LC}$. 
However, an earlier LC rise, accompanied by a brighter
luminosity at maximum even when the ejecta contain the same amount of \Nifs, is
a change orthogonal to the observed width-luminsity relation. 
Thus, differences in $\KE$ could cause an intrinsic dispersion 
in the width-luminosity relation.

The \Nifs\ distribution could also affect the LC rise time.  If HVG SNe have an
extended \Nifs\ distribution as suggested in \S \ref{sec:HVGLVG}, their LC may
peak earlier than in the case where \Nifs\ is confined in the innermost layers
because the diffusion time is shorter for the outer \Nifs. An early LC rise
would make the peak luminosity brighter even for the same \Nifs\ mass.
Consequently, SNe with similar $\Delta m_{15}$ may have different peak
luminosities in certain conditions of mixing (\eg Mazzali \& Podsiadlowski
2006, Woosley et al. 2007).

\section{CONCLUSIONS}
\label{sec:conclusions}

The outermost ejecta of Type Ia SNe are studied 
by modeling very early spectra.
Of the 6 SNe we studied, 2 (SNe 2002bo and 2002dj) are classified as HVG, 
being characterized by a high photospheric velocity and a low photospheric 
temperature. All three LVG SNe (SNe 2001el, 2003cg and 2003du) have uniform
properties and are characterized by a low photospheric velocity and a high
photospheric temperature. The properties of SN 2002er, which is classified as
HVG, are close to those of LVG, suggesting that there is continuity between the
groups. This is consistent with suggestions by Branch et al. (2006, 2007)

HVG SNe have a \ion{Si}{ii} $\lambda$6355 line with a pronounced emission peak
and a broad blue absorption, suggesting that Si is present at $v > 20,000$
\kms. The abundance distribution in HVG SNe is similar to that of a delayed 
detonation model, \ie the burning front reaches the outermost layers. On the
other hand, the \ion{Si}{ii} $\lambda$6355 emission profile of LVG SNe tends to
be suppressed, possibly because of the presence of \ion{C}{ii} $\lambda$6578.
This suggests that the burning front in LVG SNe is weaker than in HVG SNe.
However, the mass fraction of carbon at the photosphere of LVG SNe 
is only $X$(C) $\lsim 0.01$, and
the estimated C mass is less than $\sim 0.01 \Msun$, which is less 
than predicted by W7 ($\sim 0.05 \Msun$).

The difference in the photospheric velocity explains the different appearance 
of HVG and LVG SNe at pre-maximum phases through the temperature difference.
At the highest velocities ($v \gsim 20,000$ \kms), however, the variety of the 
\ion{Ca}{ii} and \ion{Si}{ii} features is not explained by this scenario.
This may require additional factors such as asphericity or diversity 
in the element abundance of the progenitor WD.

Both stable Fe and \Nifs\ are detected at the photosphere ($v = 9-14,000$ \kms)
in all 6 SNe, suggesting that Fe-group elements are always present in the 
outer layers.
The differece in their abundances may be an important reason for the 
diversity among SN Ia spectra.
On the other hand, the difference in kinetic energy between HVG and LVG SNe 
is too small to be the origin of the spectral diversity.

The diversity of SNe Ia seen in early phase spectra could cause the intrinsic
dispersion in the LC width-luminosity relation, through the difference in
kinetic energy and the amount of iron-group elements in the outer layers.

\acknowledgments
M.T. is supported by the JSPS (Japan Society for the Promotion of Science) 
Research Fellowship for Young Scientists.
G.P. acknowledges support by the Proyecto FONDECYT 3070034.
This research was supported in part by the Grant-in-Aid for 
Scientific Research (18104003, 18540231)
and the 21st Century COE Program (QUEST) from the JSPS and MEXT of Japan.

\end{document}